\documentclass[sigconf, natbib=false]{acmart}

\AtBeginDocument{%
  }

\setcopyright{acmcopyright}
\copyrightyear{2023}
\acmYear{2023}
\acmDOI{XXXXXXX.XXXXXXX}

\acmConference[CBMI]{}{September 20--22,
  2023}{Orleans, France}
\acmPrice{15.00}
\acmISBN{978-1-4503-XXXX-X/18/06}

\usepackage[backend=biber,style=acmauthoryear,maxcitenames=4]{biblatex}
\usepackage[utf8]{inputenc}
\begin{document}

\title{Deep unsupervised domain adaptation applied to the Cherenkov Telescope Array Large-Sized Telescope}

\author{Michaël Dell'aiera}
\orcid{0000-0002-5221-0240}
\email{dellaiera@lapp.in2p3.fr}
\affiliation{
  \institution{LAPP, CNRS}
  \city{Annecy}
  \country{France}
  \postcode{74940}
}

\author{Mikaël Jacquemont}
\orcid{0000-0002-4012-6930}
\email{jacquemont@lapp.in2p3.fr}
\affiliation{
  \institution{LAPP, CNRS}
  \city{Annecy}
  \country{France}
  \postcode{74940}
}

\author{Thomas Vuillaume}
\orcid{0000-0002-5686-2078}
\email{vuillaume@lapp.in2p3.fr}
\affiliation{
  \institution{LAPP, CNRS}
  \city{Annecy}
  \country{France}
  \postcode{74940}
}

\author{Alexandre Benoit}
\orcid{0000-0002-0627-4948}
\email{alexandre.benoit@univ-smb.fr}
\affiliation{
  \institution{LISTIC}
  \city{Annecy}
  \country{France}
  \postcode{74940}
}

\author{for the CTA-LST\\ Project}

\renewcommand{\shortauthors}{Dell'aiera et al.}

\begin{abstract}
  The Cherenkov Telescope Array is the next generation of observatory using imaging air Cherenkov technique for very-high-energy gamma-ray astronomy. Its first prototype telescope is operational on-site at La Palma and its data acquisitions allowed to detect known sources, study new ones, and to confirm the performance expectations. The application of deep learning for the reconstruction of the incident particle physical properties (energy, direction of arrival and type) have shown promising results when conducted on simulations. Nevertheless, its application to real observational data is challenging because deep-learning-based models can suffer from domain shifts. In the present article, we address this issue by implementing domain adaptation methods into state-of-art deep learning models for Imaging Atmospheric Cherenkov Telescopes event reconstruction to reduce the domain discrepancies, and we shed light on the gain in performance that they bring along.
\end{abstract}


\begin{teaserfigure}
  \includegraphics[width=\textwidth]{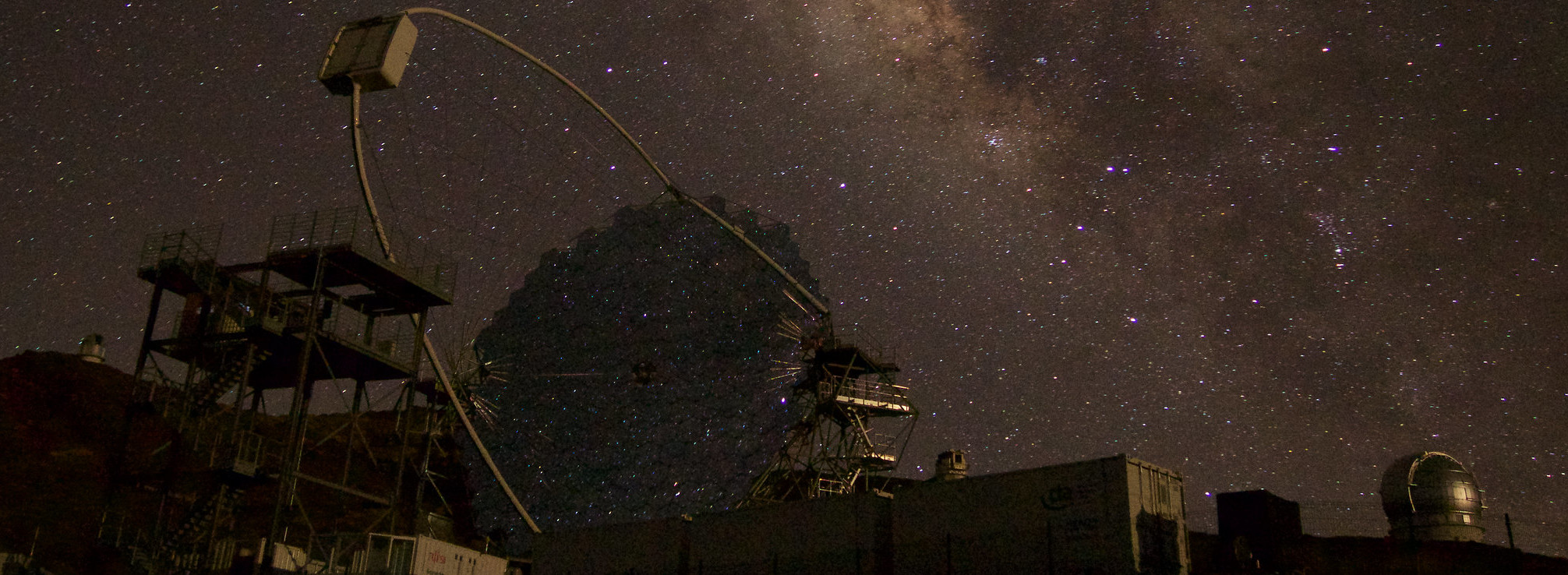}
  \caption{The LST-1 prototype. Image credit: Moritz Hütten and Dirk Hoffmann}
  \label{fig:teaser}
\end{teaserfigure}


\maketitle

\section{Introduction}
\label{sec:Introduction}

\begin{figure}
    \centering
    \includegraphics[width=\columnwidth]{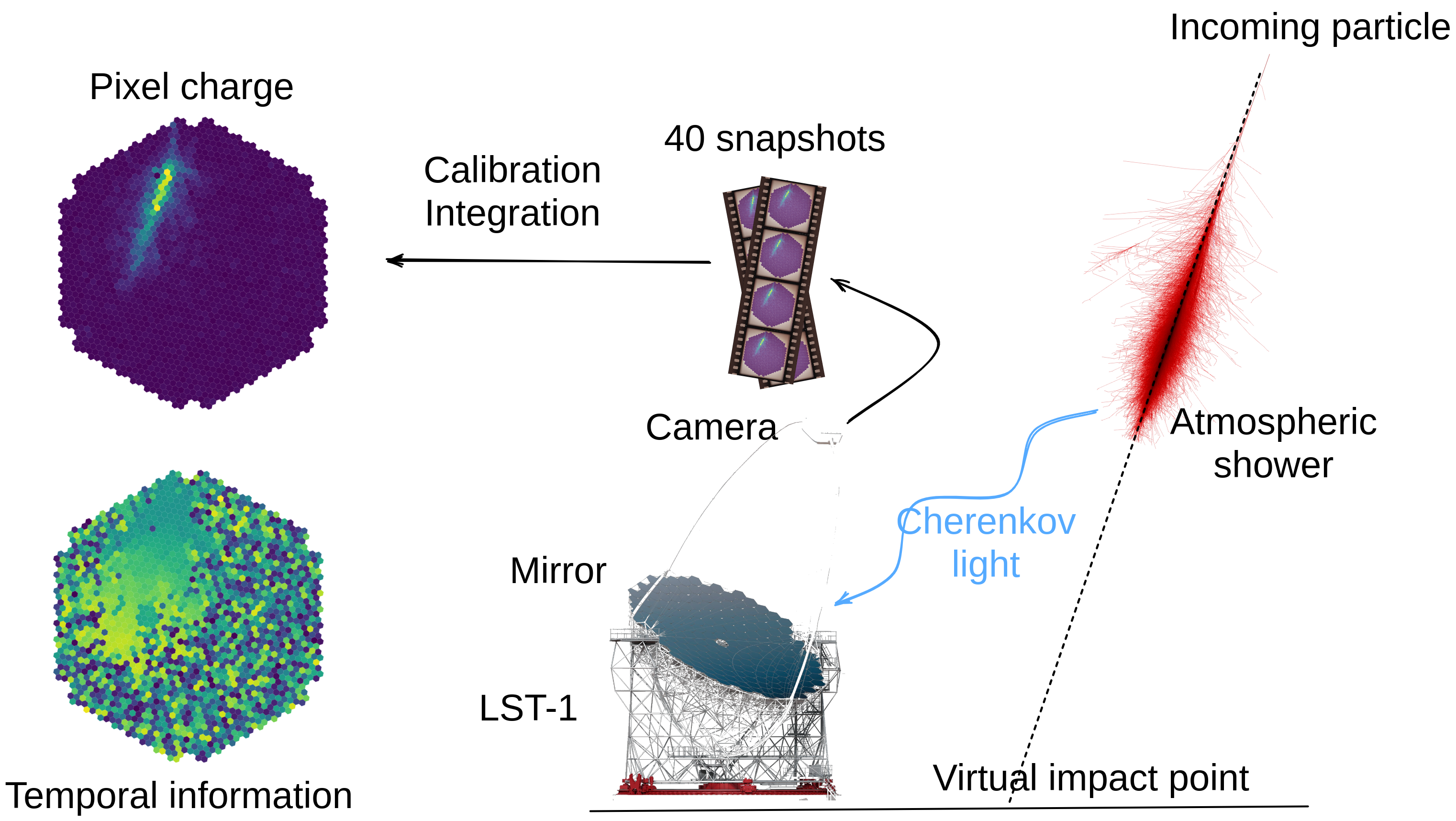}
    \caption{The working principle of IACTs is broadly a three-step procedure. First, an incident particle interacts with the atmosphere, resulting in a particle shower which emits Cherenkov photos that are focused on the telescope's camera using a mirror surface. Secondly, the camera will record a sequence of 40 snapshots within 40 ns. Lastly, this video is calibrated and integrated into two single images that are the pixel charge and the temporal information, the latter corresponding to the peak arrival time in each pixel.
    }
    \label{fig:principle_detection}
\end{figure}

Very-high-energy (VHE) gamma-ray astronomy is a field of physics that studies celestial objects and phenomena through the observation of the most energetic form of electromagnetic radiation. This discipline has made great progress in the last decades and its scientific potential covers a wide range of applications \cite{hofmann2018cta} \cite{book2018}, such as:
\begin{itemize}
    \item The understanding of the origin of cosmic particles – along with their role on star formation and galaxy evolution.
    \item The processes at stake in black holes or neutron stars' surroundings.
    \item Probing cosmology and fundamental physics, e.g the nature of dark matter \cite{Abdalla_2021}.
\end{itemize}
For this purpose, the last two decades have witnessed the birth of several major Imaging Atmospheric Cherenkov Telescopes (IACTs), namely the High Energy Stereoscopic System (H.E.S.S.) \cite{2006A&A...457..899A}, the Major Atmospheric Gamma-ray Imaging Cherenkov Telescope (MAGIC) \cite{ALEKSIC2012435}, and Very Energetic Radiation Imaging Telescope Array System (VERITAS) \cite{Weekes:2001pd}. The Cherenkov Telescope Array (CTA) aims to go beyond by improving the sensitivity by a factor of five to ten compared to the current-generation instruments and provide an energy coverage from 20 GeV to more than 300 TeV. Although the project is currently in the construction phase, the first Large-Sized Telescope (LST-1, portrayed in Figure \ref{fig:teaser}) prototype is under commissioning yet operational and has already made its first detections \cite{Abe:2021Sq,lst_lhaaso}. Similarly to all IACTs, the detection of gamma rays with CTA is not straightforward as the principle of detection is indirect and relies on using the atmosphere as a calorimeter. When a gamma ray or a charged cosmic ray interacts with the atmosphere, it triggers an extended air shower (EAS), where the particles emitted in the shower travel faster than the speed of light in this medium.. Analogically to the sonic boom emitted by a plane flying faster than the speed of sound, it results in an emission of photons from the particle shower, the Cherenkov radiation \cite{Jelley_1955}.  The Cherenkov photons are then collected by a optical system and focused onto a camera plane. After calibration and signal integration, the information of the recorded events are compressed into two images containing the pixel charge and the peak arrival times (see Figure \ref{fig:principle_detection}). Those images are then further analysed to reconstruct the incident particle physical parameters that are the energy, the direction of arrival, and its type. However, as the detected events are dominated by cosmic-ray-initiated showers, the sensitivity of IACTs is strongly dependent on the ability to distinguish between the particle types. Machine learning methods have successfully been applied for this purpose, but have seen limitations regarding lower energies, and this suggests developing new techniques to further improve the performances.

In the data analysis toolbox, artificial intelligence and more recently its deep learning subfield has drawn the attention of the scientific community due to its incredible performance on various disciplines. In particular, in the computer vision domain, convolutional neural networks (CNNs) \cite{lecun1998cnn} emerged to be a powerful tool for image data analysis. However, as it is impossible to obtain error-free labelled data, the training exclusively relies on simulations, leading in most cases to degraded results when applied to real observations \cite{jacquemont2021cta}. Fortunately, a whole set of algorithms and techniques has been developed to tackle this problem and is referred to as domain adaptation \cite{zhao2020review}. 

In this paper, we shed light on the importance of domain adaptation for CTA image analysis, and implement three methods that are incorporated into a state-of-the-art DL-based architecture for IACTs event reconstruction featuring multitask balancing and compare their respective performances. This paper is organized as follows: the standard analysis along with the related work on deep learning applied to IACTs is first introduced in Section \ref{sec:Related work}. Then, the proposed method is detailed in Section \ref{sec:Present work}, and Section \ref{sec:Results} illustrates the obtained results. Finally, Section \ref{sec:Conclusion} will draw the conclusions, and pave the way to new perspectives.

\section{Related Work}
\label{sec:Related work}

\subsection{The standard analysis and its limitations}
The standard analysis for IACTs is based on classical machine learning algorithms to extract relevant features from the images, and the application of deep learning is still in its early stages. The Hillas \cite{hillas1985hillas} method's principle is to extract the parameters of the image based on its properties and moments. These parameters are then fed to random forests \cite{ohm2009bdt} to predict the physical properties of the primary particles, and is hereby referred to as \textit{Hillas+RF}. By design, the performance of this algorithm drops at lower energies as it is significantly harder to reconstruct faint images, where the Cherenkov signal is only contained within a few number of pixels. More elaborate, template-based methods, such as Impact \cite{parsons2014impact} and Model++ \cite{denaurois2009modelpp}, match the integrated inputs with a bank of images through a likelihood function. Although these techniques are more performant, especially at lower energies, they fail at scaling to numerous telescopes as the calculation is time and resource-consuming, and will pose major challenges to be able to cope with the tremendous amount of data that CTA will provide. 
Besides, all of these techniques suffer in  different ways from the discrepancies of simulations and observational data. The Hillas+RF method cope with it thanks to an image cleaning procedure that removes most of the noises, and, in more recent works, by modifying the training data distribution with the addition of a Poisson noise to match the Night Sky Background (NSB) distribution. Although this yields to an increase in performance \cite{jacquemont2021cta}, it only takes into account a specific known difference between the domains, which can only lead to a limited gain. The Model++ is less sensitive to these differences, as NSB modelling is part of the fitting procedure.

\subsection{Deep learning applied to IACTs}
Deep learning aims at solving most of the issues mentioned previously. Using the complete image information, it is expected to improve the performances (such as sensitivity, energy resolution or angular resolution), especially at low energies. Another advantage of this approach is that it does not require any cleaning procedure to apply to the input images, on the contrary to the Hillas method. The next section describes the current state-of-the-art regarding the application of deep learning for each IACT collaboration.

H.E.S.S. is an array of five telescopes located in Namibia. During its first phase, four 12-meter diameter instruments were constructed, producing images of 960 hexagonal pixels to be analysed. Shilon et al \cite{shilon2019hess} addresses the gamma / hadron separation and the direction reconstruction of the primary particle using a combination of a CNN and a Recurrent Neural Network (RNN). Each telescope image passes through a CNN and each output feeds an RNN layer. Although the authors shed light on an improvement for the gamma/proton classification, they measured similar results compared to Hillas+RF for the direction regression. Similarly, Parsons et al \cite{parsons2020hess} also jointly employ a CNN and an RNN to approach the gamma/proton separation, but they included the Hillas parameters as an auxiliary input of the neural network. However, they concluded that their approach is sensitive to the sky brightness in the region of the observed source. De et al \cite{de2022hess} handle three distinct tasks that are the background rejection, a multi-category classification for the specific particle class categorization, and an anomaly detection to classify whether the incident particle falls into the standard model particles. They designed a CNN for the supervised context and an auto-encoder for the unsupervised one, under the assumption that the reconstruction error associated to an anomaly must increase compared to the mostly represented images in the training set. Authors concluded that their classifiers obtained state-of-the-art accuracy for background rejection, and encouraging results in the case of the multi-category classification.

MAGIC is a gamma-ray observatory composed of two telescopes located in La Palma. Authors from \cite{miener2021magic} implemented three single-task CNN-based models to regress the energy, the direction of arrival or the particle type separately. Furthermore, likewise \cite{shilon2019hess}, the stereoscopy is introduced by concatenating the images into a unique multi-channel image. The domain confusion problem is tackled by cleaning the images before training the models, but this approach relies on the same image cleaning routines as the standard analysis. Besides, authors obtained a similar sensitivity of detection regarding real acquisitions compared to the standard analysis.

CTA is the new generation of IACTs located in La Palma and Chile. Currently, the first Large-Sized Telescope (LST-1) is under commissioning but has already collected real data. Nieto et al \cite{nieto2017cta} explored the single-image classification framework using CNNs, and highlighted the feasibility of the approach. Mangano et al \cite{mangano2018cta} tackle the full-event reconstruction problem with multiple CNN, and have shown promising results. Nieto et al \cite{nieto2021cta} define a monoscopic TRN-single-tel model, embedding a shallow CNN with residual connections \cite{he2015resnet}. Although they achieve a full-event reconstruction, each task is addressed separately with a specific network. Jacquemont et al \cite{jacquemont2021cta} implement the first full-event reconstruction with the $\gamma$-PhysNet \cite{jacquemont2021thesis}. Oppositely to the Hillas+RF mono-task standard analysis, $\gamma$-PhysNet reconstructs each parameter simultaneously. However, training the weights on simulated data introduces biases when applied to real acquisitions. As a result, this real case analysis illustrates that domain adaptation has a great importance in the improvement of the $\gamma$-PhysNet.

\section{Present Work}
\label{sec:Present work}

\begin{figure}
    \centering
    \includegraphics[width=\columnwidth]{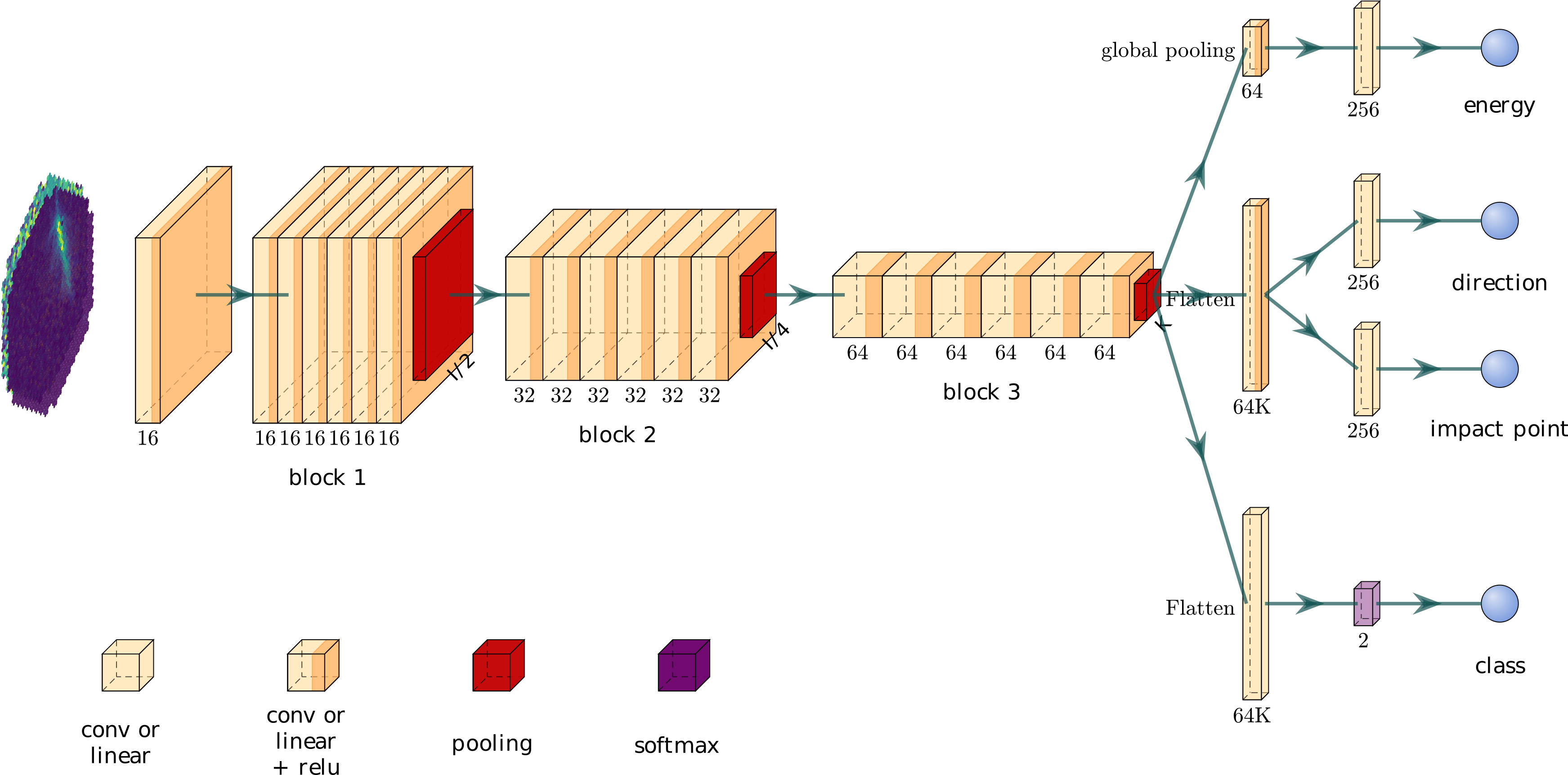}
    \caption{The $\gamma$-PhysNet architecture. The encoder used is a ResNet with attention mechanisms.}
    \label{fig:gpn}
\end{figure}

In this section, we first introduce the $\gamma$-PhysNet architecture in Figure \ref{fig:gpn}. Then, we describe the domain adaptation paradigm and the implemented methods. Finally, we demonstrate how domain adaptation can be integrated into the multi-task balancing framework.
The code implementation presented in this work is open-source and developed within the \texttt{GammaLearn}\footnote{\href{https://gitlab.in2p3.fr/gammalearn/gammalearn}{https://gitlab.in2p3.fr/gammalearn/gammalearn}} framework \cite{Jacquemont:2019pw, gammalearn_zenodo_v010}.
 
\subsection{$\gamma$-PhysNet architecture}

The $\gamma$-PhysNet has been presented in \cite{visapp21}. The network consists of two entities that are a ResNet backbone, augmented with attention mechanisms \cite{hu2019squeezeandexcitation}, and a multi-task architecture that successively decomposes the flow into the corresponding parameters to recover. The former is described as $G_f$ of parameters $\theta_f$, whereas the latter is defined as ($G_{\epsilon}$, $G_{\alpha}$, $G_{\delta}$, $G_c$) of parameters ($\theta_{\epsilon}$, $\theta_{\alpha}$, $\theta_{\delta}$, $\theta_c$) for respectively the energy, the direction, the impact point and the class. The global objective function can be implemented as the weighted sum of each component's loss as follows:

\begin{equation}
    \begin{array}{ll}
    \mathcal{L}(\theta_{f}, \theta_{\epsilon}, \theta_{\alpha}, \theta_{\delta}, \theta_{c})
         & = \mathcal{L}_{\gamma_{PN}}(\theta_{\gamma_{PN}}) \\
         & = \lambda_{\text{energy}} \sum_{i} MAE\left(G_{\epsilon}^{\theta_{\epsilon}}(G_f^{\theta_f}(x_i)), y_i\right) \\
         & + \lambda_{\text{direction}} \sum_{i} MAE\left(G_{\alpha}^{\theta_{\alpha}}(G_f^{\theta_f}(x_i)), y_i\right) \\
         & + \lambda_{\text{impact}} \sum_{i} MAE\left(G_{\delta}^{\theta_{\delta}}(G_f^{\theta_f}(x_i)), y_i\right) \\ 
         & + \lambda_{\text{class}} \sum_{i} CE\left(G_c^{\theta_c}(G_f^{\theta_f}(x_i)), y_i\right)
    \end{array}
\end{equation}

\noindent
where $MAE$ and $CE$ respectively designate Mean Absolute Error and Cross-Entropy while tuple $(x_i, y_i)$ correspond to the input and corresponding classification label or regression target values. This basis is used to further describe the domain adaptation procedure.

\subsection{Domain adaptation}

As previously discussed in the Section \ref{sec:Related work}, the performance is limited by the intrinsic differences between the training simulations and the real observations. There are many possible sources of discrepancies, and the following list is not exhaustive:
\begin{itemize}
    \item The theoretical models underlying the simulations and their production approximate the reality closely, but unknown physics interactions and computing simplifications occur.
    \item The Night Sky Background (NSB) distribution, which corresponds to the background light in the night sky, depends on the observed sky region, and varies in time. For example, stars in the field of view will ultimately disturb the inference if they are not included in the training dataset.
    \item Models are trained on simulations, but the training data sparsely cover the parameter space (for example the telescope pointing direction which is simulated following a predefined grid). 
\end{itemize}

The analysis strategy that consists in adding a Poisson noise to the training data is hard to adapt run by run. Fortunately, domain adaptation is a wide and dynamic field of deep learning that copes with distribution discrepancies between a source and a target domain. More specifically, in the case of IACTs where obtaining error-free source labels is practically impossible, the methodology is constrained to its subfield called deep unsupervised domain adaptation. Although real observations are unlabelled, it is possible to artificially generate a binary domain label to each data representing its affiliation to the simulations or the real acquisitions.

\subsection{Implemented methods}

We selected three popular unsupervised domain adaptation methods, namely Domain Adversarial Neural Network (DANN) \cite{ganin2015dann}, Deep Joint Distribution Optimal Transport (DeepJDOT) \cite{courty2018deepjdot} and Deep Correlation Alignment (DeepCORAL) \cite{sun2016deepcoral}. 

Firstly, DANN extends any neural network with the addition of a domain classifier $G^d$ of parameters $\theta_d$ in parallel to the classification and regression branches. It consists in confronting the feature extractor and the domain classifier in an adversarial way so that the backbone learns a domain-invariant representation of the data.The adversarial approach is implemented using the Gradient Reversal Layer (GRL) $\mathcal{R}$ defined as:
\begin{equation}
    \mathcal{R}(x) = x \ \ \text{ and } \ \ \dfrac{d\mathcal{R}}{dx} = -K \times I
\end{equation}
\vspace{-0.2cm}
where $K$ can be any constant or time-dependent variable and $I$ refers to the identity matrix. Thus, the global objective function can be computed as follow:
\begin{equation}
    \begin{array}{ll}
    \mathcal{L}(\theta_{\gamma_{PN}}, \theta_d)
         & = \mathcal{L}_{\gamma_{PN}}(\theta_{\gamma_{PN}}) \\
         & - \lambda_{\text{domain}} \sum_{i} CE\left(G_d^{\theta_d}(\mathcal{R}\left[G_f^{\theta_f}(x_i)\right]), d_i\right)
    \end{array}
\end{equation}

Secondly, the fundamental concept of DeepJDOT relies on the optimal transport theory to compute the Wasserstein metric which, when minimized, increases domain confusion. This metric can be computed using an optimal transport plan $\pi$ and a cost matrix $C$ as defined in \cite{courty2018deepjdot}. This method can be easily integrated into any framework as it only consists in adding an extra Wasserstein loss to the global objective function:
\begin{equation}
    \begin{array}{ll}
        \mathcal{L}(\theta_{\gamma_{PN}}) & = \mathcal{L}_{\gamma_{PN}}(\theta_{\gamma_{PN}}) \\
         & + \lambda_{\text{domain}} \sum_{i,j} \pi_{i,j} C_{i,j} \\
    \end{array}
\end{equation}

Lastly, the basic idea behind DeepCORAL is to align the features correlation $C_S$ and $C_T$ respectively between the source and target domains through the addition of a coral loss to the objective function:
\begin{equation}
    \begin{array}{ll}
        \mathcal{L}(\theta_{\gamma_{PN}}) & = \mathcal{L}_{\gamma_{PN}}(\theta_{\gamma_{PN}}) \\
         & + \lambda_{\text{domain}} \dfrac{1}{4d^2} ||C_S - C_T||^2_F \\
    \end{array}
\end{equation}
where $||.||^2_F$ denotes the Frobenius norm and $d$ refers to the size of the feature vector.

To summarize, for each method the global loss is computed as the sum of $\gamma$-PhysNet model and  domain adaptation (DAD) losses:
\begin{equation}
    \mathcal{L} = \mathcal{L}_{\gamma_{PN}} + \lambda_{\text{domain}} \mathcal{L}_{\gamma_{DAD}}
\end{equation}

\subsection{Domain adaptation in the context of multi-task balancing}

Multi-task learning \cite{caruana2004multitask} is defined as the concurrent optimization of different but related tasks $t$, so their training is gainful for the others \cite{he2017maskrcnn}. In a traditional learning framework, the global objective function of a model is designed as a weighted sum of the task-specific losses $L_t$:

\begin{equation}\label{eq:mt}
    \mathcal{L} = \sum_t \lambda_t L_t
\end{equation}

However, despite being crucial to ensure a well-behaved convergence, computation $\lambda_t$ weights is often empirical. In order to mitigate costly manual optimization procedures, authors of \cite{kendall2018multitask} use the task-dependent uncertainty - also called homoscedastic uncertainty - to automatically calculate weights. 
²
\begin{equation}
    \mathcal{L}(s_1, ..., s_N) = \sum_{t=1}^{N} \dfrac{1}{2} e^{-s_t} L_t + s_t
\end{equation}
with $s_t = \text{log } \sigma_t^2$ the log-variance of the task $t$. This strategy has been implemented in \cite{jacquemont2021cta}, and yields a better performance than compared to manually tuning the weights. Moreover, the introduction of an auxiliary task - the impact point regression - helps constrain the problem, and improves the full-event reconstruction. Including domain adaptation into the multi-task balancing framework aims at computing each weight $\lambda_{\text{energy}}$, $\lambda_{\text{direction}}$, $\lambda_{\text{impact}}$, $\lambda_{\text{class}}$ and $\lambda_{\text{domain}}$ as a tuple $(0.5\times e^{-s_t}, s_t)$, and their corresponding log-variances are set as trainable parameters of an additional model. Besides, for DANN, the GRL constant $K$ is set to $K = 1$.

\section{Results}
\label{sec:Results}

In this section, we introduce the model's training data, and the figures of merit that allow comparison between the different methods. Then, we present the results obtained with our methods.

\subsection{Datasets}

\begin{figure*}
    \centering
    \includegraphics[width=\linewidth]{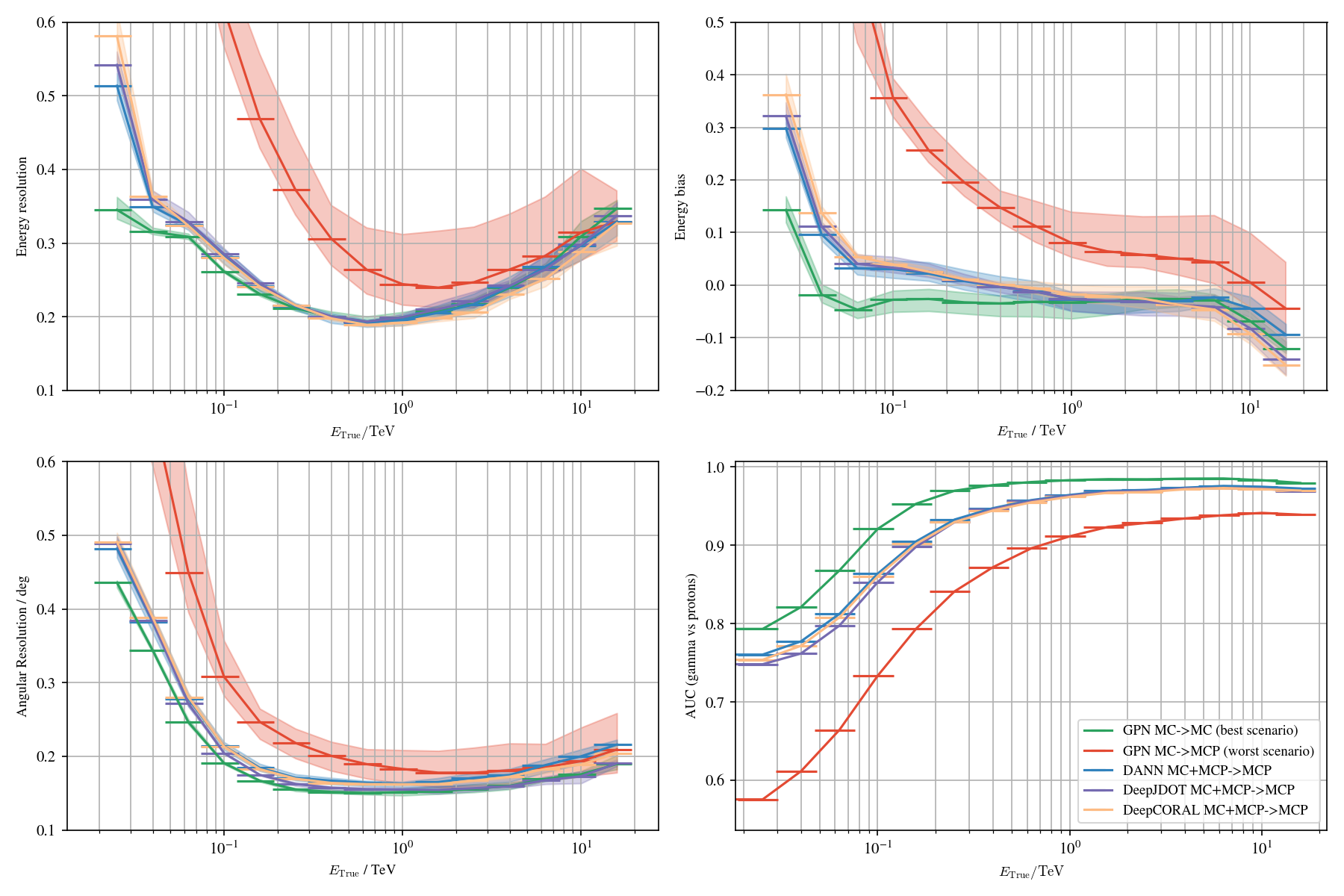}
    \caption{The IRFs obtained from the experiments using $\gamma$-PhysNet (GPN), DANN, DeepJDOT and DeepCORAL. They are averaged on five seeds. The min and the max are displayed as a surface area. Top left is the energy resolution (lower is better). Top right is the energy bias (lower is better). Bottom left is the angular resolution (lower is better). Bottom right is the AUC per energy bin (high is better). All metrics are function of the gamma-ray true energy. For the energy and angular resolutions, the lower the better whereas for the AUC, the higher the better. An energy bias equals to zero is best.}
    \label{fig:irfs}
\end{figure*}

In this work, we use the LST project dataset of Monte-Carlo simulations generated with CORSIKA \cite{heck1998corsika} and\\ \texttt{sim\_telarray} \cite{BERNLOHR2008149}, and referred to as Prod5 LST-1 mono-trigger, with a telescope response for a pointing zenith angle of $20^{\circ}$ and azimuth $180^{\circ}$. This dataset is split into a train and a test set, and $20\%$ of the training set is reserved for the validation. The training set contains 1.1M diffused gammas and 0.76M protons whereas the test set contains 1.1M point-source gammas mimicking a gamma-ray point source and 0.76M protons. The training labelled data will be referred to as source and the unlabelled data as target.

\subsection{Figures of merit}

In the case of IACTs, the performance metrics used to evaluate the models performances correspond to the Instrument Response Functions (IRFs) and are calculated with the test labelleled simulations. The expected performances for CTA are presented on the CTA webpage\footnote{\url{https://www.cta-observatory.org/science/ctao-performance/}}. We summarize below the ones used in this work.


\paragraph{Effective collection area}
The effective collection area is computed as the number of particles identified as gammas divided by the number of simulated gammas times the simulated ground area. In the present work, this metrics is fixed to compare each experiment fairly by selecting reconstructed gammas with a gammaness (the probability of an event to be a gamma as output of the network) higher than the threshold necessary to keep 70\% of them in each energy bin. As each experiment has by construction the same effective area, it is not presented in Figure \ref{fig:irfs}.

\paragraph{Energy resolution and bias}

The energy resolution and bias are respectively defined as the interval containing 68\% and the median of the difference between the gamma-ray simulated and reconstructed energy.

\paragraph{Angular resolution}

The angular resolution is computed as the interval containing 68\% of the distribution of the angular distance between the gamma-ray simulated and reconstructed direction in the sky.

\paragraph{Gamma classification}
The Area Under the ROC Curve (AUC) is defined as the integral of the Receiver Operating Characteristic (ROC) function of the gamma/proton classifier. Here we show AUC as a function of the gammas true energy to evaluate the classifier discrimination power. The higher the AUC score, the better the gamma/hadron classification, where a perfect classifier would have an AUC value of 1 and a random classifier of 0.5.

\subsection{Training parameters}
Each model is trained with 50 epochs, and corresponds to the $\gamma$-PhysNet developed in \cite{jacquemont2021thesis}. For DANN, the domain classifier is made of two fully-connected layers of 100 features, and the hyperparameter $\gamma$ is set to $10$. The Adam optimizer with a learning rate of $1e^{-3}$ is used to update the models, and the batch size is 256 for both domains. Each experiment is averaged on five different seeds. The second model calculating the loss weighting coefficients uses SGD with a learning rate of $1e^{-4}$. In all cases, the optimizers have a weight decay of $1e^{-4}$. The raw hexagonal input images are interpolated on a regular grid of size 55x55 using a bilinear interpolation. In total, $\gamma$-PhysNet, DeepJDOT and DeepCORAL have $3.5M$ parameters, whereas DANN has $4.8M$.

\subsection{Performance on simulated data}

In this work, we use standard Monte Carlo simulations as the source dataset. The target and the test datasets are composed of biased simulations and are denoted as MCP. The added noise follows a Poisson distribution with $\lambda=0.4$, adopting the approach of \cite{jacquemont2021cta}. During the model's training, the labels are removed from the target dataset. The validity of the approach is demonstrated by testing the network on biased but labelled data, and therefore get a good insight on the methods performance. The same approach could be used by replacing the target and test data with real data from LST-1, though the class balance between the source and the target datasets will be different.

The results are presented in the Figure \ref{fig:irfs}. For each experiment, models are trained with five different seeds to show the variability of the results with respect to model initialization. In order to evaluate the contribution of domain adaptation, we define a best and a worst scenario that correspond to the lower and upper performance bounds. More precisely, the best scenario describes the training of the $\gamma$-PhysNet on Monte Carlo (MC), and the inference on the test data following the same distribution (\textit{MC->MC}). Training and testing the network on biased data (\textit{MCP->MCP}) gives identical IRFs compared to the best scenario. Conversely, the worst scenario is defined by a training on Monte Carlo but tested on MCP (\textit{MC->MCP}). 

As shown here, the worst scenario displays degraded performance compared to the best scenario along with an increase of variability with respect to model initialization and random processes (e.g. shuffling). This is expected as the train and test data have different distributions, and therefore the model suffers from a significant reconstruction bias.

The three domain adaptation techniques tested here show an improvement in performance compared to the worst scenario, sometimes even matching the best scenario (in particular for the angular and energy resolutions). It can be noted that the reconstruction tasks become harder at the lowest energies as the signal/noise ratio is lower, and therefore the impact of the added noise is more important, and cannot be entirely overcome by the domain adaptation. The impact on performances for the regression tasks can be particularly noted below 200 GeV where the domain adaptation models reconstruction cannot match the best scenario.
The classification task seems to suffer the most from the addition of noise, and that the best scenario performance cannot be matched, even at the highest energies. 

Computationally wise, it can be noted that the calculation complexity of the domain adaptation methods does not significantly differ compared to the original $\gamma$-PhysNet, but the use of extra target data (in the same amount as source data) increases the training time.

\section{Conclusion and Perspectives}
\label{sec:Conclusion}

In this paper, we successfully applied deep unsupervised domain adaptation to LST-1 Monte-Carlo simulated data. We have shown that it can help the network overcome a stronger NSB in the test dataset. Our approach results in performance close to the best scenario, meaning that it successfully corrected the distribution bias introduced by the addition of the Poisson noise. Furthermore, the inclusion of domain adaptation into the multi-task balancing ease the tuning of the hyperparameters, thus it is time and computationally more efficient. The framework developed here will be tested on real data to solve the domain shift between the Monte-Carlo simulations and the telescope recorded data in a subsequent work.

\section{Acknowledgements}

We gratefully acknowledge financial support from the agencies and organizations listed here:\\\url{https://purl.org/gammalearn/acknowledgements} and feedback from CTA members. 

\clearpage


\printbibliography

\end{document}